\documentclass[prb,twocolumn,showpacs]{revtex4}

\usepackage{graphicx}

\begin{document}

\title{Interaction-enhanced electron-hole and valley asymmetries 
\\ in the lowest Landau level 
of $ABA$-stacked trilayer graphene}
\author{K. Shizuya}
\affiliation{Yukawa Institute for Theoretical Physics\\
Kyoto University,~Kyoto 606-8502,~Japan }

\begin{abstract}
In a magnetic field graphene trilayers  support a special multiplet 
of 12 zero(-energy)-mode Landau levels 
with a threefold degeneracy in Landau orbitals. 
A close look is made into such zero-mode levels
in $ABA$-stacked trilayers, 
with the Coulomb interaction taken into account.
It turns out that
the zero-mode Landau levels of $ABA$ trilayers are greatly
afflicted with electron-hole and valley asymmetries,
which come from general hopping parameters and 
which are enhanced by the Coulomb interaction 
and the associated vacuum effect, the orbital Lamb shift, 
that lifts the zero-mode degeneracy.
These asymmetries substantially affect 
the way the zero-mode levels evolve, with filling, via Coulomb interactions;
and its consequences are discussed in the light of experiments. 

\end{abstract} 


\maketitle

\section{Introduction}

Graphene, 
an atomic layer of graphite that supports massless Dirac fermions, 
displays remarkable and promising electronic properties.
Recently there is increasing interest in 
bilayers and few layers of graphene,
where the physics and applications of graphene become richer,
with, e.g., a tunable band gap~\cite{MF,OBSHR,Mc,CNMPL}
for bilayer graphene.

There are some key signatures of Dirac fermions that
distinguish graphene from conventional electron systems.
(i) In a magnetic field, graphene supports,
as  the lowest Landau level (LLL),
a special set of four zero-energy levels differing in spin and valley, 
as observed via the half-integer quantum Hall effect.
(ii)  Graphene is an intrinsically many-body system
equipped with 
the valence band acting as the Dirac sea.
Quantum fluctuations of the filled valence band are fierce,
even leading to ultraviolet divergences;
and one encounters such many-body phenomena 
as velocity renormalization,~\cite{velrenorm}
screening of charge,~\cite{KSbgr} 
and nontrivial Coulombic corrections to 
cyclotron resonance.~\cite{JHT,IWFB,BMgr,KCC,KScr}

In multilayer graphene the zero-mode Landau levels acquire a new aspect.
Bilayer graphene supports eight such levels, 
with an extra twofold degeneracy~\cite{MF} 
in Landau orbitals $n$=0 and 1.  Trilayer graphene has 12 such levels 
with threefold $\lq\lq$orbital" degeneracy, and so on. 
This orbital degeneracy is a new feature 
peculiar to the LLL in multilayer graphene, 
and leads to intriguing 
quantum phenomena~\cite{BCNM,KSpzm,BCLM,CLBM,CLPBM,CFL} 
such as orbital mixing and orbital-pseudospin waves.  
In real samples these zero-energy levels evolve, due to general interactions,
into a variety of pseudo-zero-mode (PZM) levels, 
or broken-symmetry states within the LLL, 
as discussed theoretically.~\cite{BCNM, NL}

It has been unnoticed until recently 
that many-body effects work to lift orbital degeneracy. 
Each zero-mode level, subjected to quantum fluctuations 
of the valence band, gets shifted differently 
within the LLL,
just like the Lamb shift~\cite{Lambshift} in the hydrogen atom.
This orbital Lamb shift was first noted~\cite{KSLs} for bilayer graphene 
and is also realized in an analogous fashion~\cite{KSLsTL}
in rhombohedral ($ABC$-stacked) trilayer graphene, 
which is a $\lq\lq$chiral" trilayer generalization of 
bilayer graphene.
This orbital shift is considerably larger in scale 
than intrinsic spin or valley breaking, 
and one has to take it into account  
in clarifying the fine structure of the LLL in multilayers.

Graphene trilayers attracted theorists' attention~\cite{GCP,KA,NCGP, KM79} 
even before experiments, and 
it has been verified experimentally~\cite{BZZL,KEPF,TWTJ,LLMC,BJVL,ZZCK,JCST} 
that the electronic properties of graphene trilayers
strongly depend on the stacking order, 
with $ABC$-stacked trilayers 
exhibiting a tunable band gap
and Bernal $(ABA)$-stacked trilayers, the most common type of trilayers, 
remaining metallic. 
Currently trilayers are under active study 
both experimentally~\cite{HNE,LVTZ} 
and  theoretically.~\cite{KMc81,KMc83,ZSMM,YRK,ZTM}

The purpose of this paper is to examine 
the orbital Lamb shift and its consequences 
in $ABA$-stacked trilayers,
with focus on electron-hole and valley asymmetries 
due to general band parameters.
It turns out that $ABA$ trilayers critically differ
in zero-mode characteristics from $ABC$ trilayers. 
In particular, the way the Coulomb interaction acts within the LLL
substantially differs between the two types of trilayers, 
leading to distinct basic filling-factor steps 
in which large level gaps appear in each of them.
In addition, the LLL of $ABA$ trilayers, 
unlike  that of $ABC$ trilayers, is greatly afflicted 
with interaction-enhanced electron-hole and valley asymmetries, 
which affect the sequence of broken-symmetry states within the LLL, 
observable via the quantum Hall effect.

In Sec.~II we examine the one-body spectrum of the PZM levels 
in $ABA$-trilayer graphene, 
and in Sec.~III  show how the orbital Lamb shift modifies their full spectrum.
In Sec.~IV  we discuss how the level spectra evolve, 
with filling, via the Coulomb interaction. 
Section~VI is devoted to a summary and discussion on 
how $ABA$ trilayers differ in zero-mode characteristics 
from $ABC$ trilayers.

\section{$ABA$-stacked trilayer graphene}  

$ABA$-stacked trilayer graphene consists of three graphene layers 
with vertically-arranged dimer bonds  $(B_{1}, A_{2})$ and $(A_{2}, B_{3})$,
where $(A_{i}, B_{i})$ 
denote inequivalent lattice sites in the $i$-th layer.
The interlayer coupling $\gamma_{0}\equiv \gamma_{B_{i}A_{i}} \sim 3$~eV 
is related to the Fermi velocity 
$v = (\sqrt{3}/2) a_{L}\gamma_{0}/\hbar \sim 10^6$ m/s in monolayer graphene.
Interlayer hopping via the nearest-neighbor dimer coupling 
$\gamma_{1}\equiv \gamma_{B_{1}A_{2}} 
=\gamma_{A_{2}B_{3}} \sim $ 0.4~eV
leads to linear (monolayer-like) and 
quadratic  (bilayer-like) spectra~\cite{GCP,KA} 
$\propto  |{\bf p}|,  {\bf p}^2$ 
in the low-energy branches $|\epsilon|< \gamma_{1}$.

The effective Hamiltonian for $ABA$-stacked trilayer graphene 
with general  intra- and interlayer couplings 
is written as~\cite{KMc83}
\begin{eqnarray}
H^{\rm tri} &=&\!\! \int\! d^{2}{\bf x}\, 
\Big[ (\Psi^{K})^{\dag}\, {\cal H}_{K} \Psi^{K}
+ (\Psi^{K'})^{\dag}\, {\cal H}_{K'}\, \Psi^{K'}\Big], \nonumber\\
{\cal H}_{K} &=& \left(
\begin{array}{lll}
D &  V &  W \\
 V^{\dag} & D & V^{\dag}\\
W & V & D \\
\end{array}
\right) + U,\ 
D = \left(
\begin{array}{cc}
0 &  v\, p^{\dag} \\
 v\, p & 0 \\
\end{array}
\right),
\nonumber\\
V &=& \left(
\begin{array}{cc}
-v_{4}\, p^{\dag}&  v_{3}\, p \\
 \gamma_{1} & -v_{4}\, p^{\dag} \\
\end{array}
\right),\ 
W= \left(
\begin{array}{cc}
\gamma_{2}/2& 0 \\
0 &\gamma_{5}/2\\
\end{array}
\right),\ \ \ 
\nonumber\\
U&=& {\rm diag}(U_{1}, U_{1} + \Delta', U_{2}+ \Delta', 
U_{2} , U_{3}, U_{3} + \Delta'),
\label{Htri}
\end{eqnarray}
with 
$p= p_{x} + i p_{y}$ and $p^{\dag}= p_{x} - i p_{y}$.
Here  $\Psi^{K}=(\psi_{A_{1}}, \psi_{B_{1}}, \psi_{A_{2}},\psi_{B_{2}},
\psi_{A_{3}}, \psi_{B_{3}})^{\rm t}$
stands for the electron field at the $K$ valley.
$v_{3}$ and $v_{4}$ are related to the nonleading nearest-layer coupling
$\gamma_{3} \equiv \gamma_{A_{1} B_{2}}$ and
$\gamma_{4} \equiv \gamma_{A_{1} A_{2}}= \gamma_{B_{1} B_{2}}$, 
respectively.
$\gamma_{2} \equiv \gamma_{A_{1}A_{3}}$ and 
$\gamma_{5} \equiv \gamma_{B_{1}B_{3}}$ describe coupling 
between the top and bottom layers.
$(U_{1}, U_{2}, U_{3})$ denote the on-site energies of the three layers;
we take $U_{2}=0$ without loss of generality 
and focus on the case of a symmetric bias~\cite{GCP} 
$U_{3}= -U_{1} \equiv u$.
Such an interlayer bias leads to a tunable band gap 
for $ABC$-stacked trilayers, but not for $ABA$-stacked trilayers 
which involve monolayer-like subbands. 
$\Delta'$ stands for the energy difference 
between the dimer and non-dimer sites.
${\cal H}_{K}$ is diagonal in (suppressed) electron spin.

The Hamiltonian ${\cal H}_{K'}$ at another valley is given by ${\cal H}_{K}$ 
with $p \rightarrow  - p_{x} + i p_{y} = -p^{\dag}$
and $p^{\dag}\rightarrow - p$,
and acts on a spinor of the same sublattice content as $\Psi^{K}$.
${\cal H}_{K'}$ is  not linked to ${\cal H}_{K}$ in a simple way 
and, in a magnetic field, 
their Landau-level spectra significantly differ,~\cite{KMc81} 
especially for zero-mode levels,
although they precisely but nontrivially~\cite{KMc81}  agree
when only the leading parameters $(v, \gamma_{1})$ 
are kept. 
This is in sharp contrast to the case of bilayers and $ABC$-stacked trilayers,
for which ${\cal H}_{K'}$ is linked to ${\cal H}_{K}$ 
via unitary equivalence,~\cite{KSLs,KSLsTL}
such as ${\cal H}_{K'}^{ABC}  \sim {\cal H}_{K}^{ABC}|_{-v_{3}, -\gamma_{2}; 
U_{1} \leftrightarrow U_{3} }$.

For the trilayer  hopping parameters 
one may use, as typical values, those for graphite,~\cite{KMc83}
\begin{eqnarray}
&&\gamma_{0}\approx 3.16\, {\rm eV\  or} \  v\approx 1.0 \times 10^6 {\rm m/s},
\nonumber\\
&&\gamma_{1}\approx 0.4\,  {\rm  eV},
\gamma_{3}\approx 0.3\, {\rm eV},
\gamma_{4}\approx 0.04\, {\rm eV},
\nonumber\\
&&\gamma_{2}\approx -0.02\, {\rm eV}, 
\gamma_{5}\approx 0.04\, {\rm eV},
\Delta' \approx 0.05\, {\rm eV}.
\label{Triparameter}
\end{eqnarray}
In the present analysis 
we regard $(v, \gamma_{1})$ as the basic parameters
and treat the nonleading ones 
$(\gamma_{2}, \gamma_{5}, \gamma_{4}, \cdots)$ 
and bias $u$ as perturbations. 
We ignore $v_{3}\propto \gamma_{3}$ from the start 
since its effect is negligible in high magnetic fields, 
as discussed later in this section.

Let us place trilayer graphene in a strong uniform magnetic field 
$B_{z} = B>0$ normal to the sample plane;
we set, in ${\cal H}_{K}$, 
$p\rightarrow \Pi = p + eA$
with $A= A_{x}+ iA_{y}= -B\, y$, 
and scale $a \equiv \sqrt{2eB}\,  \Pi^{\dag}$ so that $[a,a^{\dag}]=1$.
It is easily seen that the eigenmodes of ${\cal H}_{K}$ have the structure 
\begin{eqnarray}
\Psi_{n} &=& \Big( |n -2 \rangle\, b_{n}^{(1)} ,|n-1 \rangle\, d_{n}^{(1)} ,
|n -1 \rangle\, b_{n}^{(2)} , \nonumber\\
&&
|n \rangle\,  d_{n}^{(2)} ,  |n -2 \rangle\, b_{n}^{(3)} , 
|n-1 \rangle\, d_{n}^{(3)}  \Big)^{\rm t}
\label{Psi_n}
\end{eqnarray} 
with $n=0,1,2,\cdots$, where only the orbital eigenmodes are shown 
using the standard harmonic-oscillator basis $\{ |n\rangle \}$
(with the understanding that $|n\rangle =0$ for $n<0$).
The coefficients 
${\bf v}_{n}=(b_{n}^{(1)}, d_{n}^{(1)}, b_{n}^{(2)}, d_{n}^{(2)}, 
b_{n}^{(3)}, d_{n}^{(3)} )^{\rm t}$
are given by the eigenvectors 
 (chosen to form an orthonormal basis)  of the reduced Hamiltonian
$\hat{\cal H}_{\rm red} \equiv \omega_{c} {\cal H}_{n}$ 
with
\begin{equation}
{\cal H}_{n} 
\!=\! \!
\left( \!\!\!
\begin{array}{cccccc}
-\hat{u} & r_{n-1} &\! - \lambda r_{n-1}&0 &R_{2}&0 \\
r_{n-1} &\! \delta -\hat{u}  &\hat{\gamma} & - \lambda r_{n} &0 &R_{5} \\
-\lambda r_{n-1} & \hat{\gamma} & \delta& r_{n} &\! - \lambda r_{n-1} & \hat{\gamma} \\
0&-\lambda r_{n}  & r_{n}  & 0 & 0 & -\lambda r_{n} \\
R_{2}& 0 &\! - \lambda r_{n-1} & 0 & \hat{u} & r_{n-1} \\
0& R_{5} & \hat{\gamma} & - \lambda r_{n} & r_{n-1} & \delta + \hat{u} \\
\end{array} \! \! 
\right),
\nonumber\\
\label{Hn}
\end{equation}
where $r_{n}\equiv \sqrt{n}$ for short; $ \hat{u} \equiv  u/\omega_{c}$,
$\hat{\gamma} \equiv \gamma_{1}/\omega_{c}$.
$\lambda \equiv \gamma_{4}/\gamma_{0}\,  (\approx 0.013)$, 
$R_{2} \equiv  (\gamma_{2}/2)/\omega_{c}$, 
$R_{5} \equiv  (\gamma_{5}/2)/\omega_{c}$ 
and $\delta\equiv \Delta'/\omega_{c}$.
Here
\begin{equation}
\omega_{c}\equiv \sqrt{2}\, v/\ell 
\approx 36.3 \times v[10^{6}{\rm m/s}]\, \sqrt{B[{\rm T}]}\ {\rm meV}
\end{equation} 
stands for the characteristic cyclotron energy 
for monolayer graphene,
with $v$ in units of $10^{6}$m/s and $B$ in teslas;
$\ell \equiv 1/\sqrt{eB}$ denotes the magnetic length.
Note that eigenvectors ${\bf v}_{n}$ can be taken real since 
${\cal H}_{n}$ is a real symmetric matrix.

Solving the secular equation shows that there are 6 branches of 
Landau levels for each integer $n\ge 2$,
with two branches of monolayer-like spectra 
$\epsilon \sim \pm \sqrt{n-1}\, \omega_{c}$
and four branches of bilayer-like spectra.
We denote the eigenvalues as
$\epsilon_{-n''} < \epsilon_{-n'} < \epsilon_{-n} < 0 < \epsilon_{n}<\epsilon_{n'}
< \epsilon_{n''}$,
so that the index $\pm n$ reflects the sign of $\epsilon_{n}$.
The $|n|=2$ levels, e.g.,  consist of the $n=(\pm 2. \pm 2',\pm 2'')$ branches.

As verified easily, with only $(v,\gamma_{1})$ and bias $u$ kept,
the spectrum and eigenvectors of 
$\hat{\cal H}_{\rm red}$ have the property
\begin{equation}
\epsilon_{- n} = - \epsilon_{n}|_{-u}, 
b_{- n}^{(i)} = - b_{n}^{(i)}|_{-u}, d_{- n}^{(i)} = d_{n}^{(i)}|_{-u}, 
\label{vn}
\end{equation}
for $|n|\ge 2$ [and each branch $(n,n',n'')$],
where $b_{n}^{(i)}|_{-u}$, e.g., stands for 
$b_{n}^{(i)}$ with $u\rightarrow -u$.
This structure~\cite{fnone}  is also seen from the fact that 
$-{\cal H}_{K}$ is unitarily equivalent to ${\cal H}_{K}$ 
with the signs of 
$(U_{i}, v_{4}, \gamma_{2},\gamma_{5}, \Delta')$ reversed, 
\begin{equation}
\Sigma_{3}^{\dag}{\cal H}_{K}\Sigma_{3}  
= - {\cal H}_{K}|_{-U_{i}, -v_{4}, -\gamma_{2}, -\gamma_{5},-\Delta'} ,
\label{Hequi} 
\end{equation}
where $\Sigma_{3} = {\rm diag} (\sigma_{3}, \sigma_{3}, \sigma_{3})$;
thus Eq.~(\ref{vn}) is generalized to the full spectrum accordingly.

There are three zero-energy solutions (per spin) 
within the $n\in (0,1)$ sector for $u\rightarrow 0$.
As seen from Eq.~(\ref{Psi_n}), for $n=0$, 
$\hat{\cal H}_{\rm red}$ is reduced to a matrix of rank 1,
with an obvious eigenvalue 
\begin{equation}
\epsilon_{0}=U_{2}= 0
\label{zmK}
\end{equation}
and the eigenvector
${\bf v}_{0} = (0,0,0,1,0,0)^{\rm t}$
or 
\begin{equation}
\Psi_{0} = ( 0,0,0, |0\rangle, 0, 0)^{\rm t}.
\label{psizero}
\end{equation}

For $n=1$, $\hat{\cal H}_{\rm red}$ has rank 4, and 
we specify the four eigenmodes 
as $n=1_{\pm}$ and $n=\pm 1'$,
with energy spectra
$\epsilon_{1_{\pm}} =\pm \sigma\, u$ and 
$\epsilon_{\pm 1'} = \pm (1/\sigma)\, \omega_{c} \sim \pm \sqrt{2}\, \gamma_{1}$
when only $(v, \gamma_{1}, u)$ are kept,
where
\begin{equation}
\sigma \approx 1/\sqrt{2\, \hat{\gamma}^2 +1} \ <1; 
\end{equation}
$\hat{\gamma} \approx 2.4$ and $\sigma \approx 0.28$ at $B=20\, $T 
with $\gamma_{1} \approx 0.4\, $eV.
For $u\rightarrow +0$, in particular, the $n=1_{\pm}$ modes have zero energy
with wave functions 
\begin{eqnarray}
\Psi_{1_{+}}^{(0)} \!\! &\stackrel{u\rightarrow 0}{=}& \! 
\Big( 0,  \alpha^{-} |0\rangle,  0,
c_{1}\, |1\rangle, 0, - \alpha^{+} |0\rangle \Big)^{\rm t},  \nonumber\\
\Psi_{1_{-}}^{(0)} \!\! &\stackrel{u\rightarrow 0}{=}& \! 
\Big( 0,  - \alpha^{+} |0\rangle,  0,
c_{1}\, |1\rangle, 0, \alpha^{-} |0\rangle \Big)^{\rm t},\ \ \ 
 \end{eqnarray}
where $\alpha^{\pm} \equiv (1\pm \sigma)/2\sim 1/2$ and
$c_{1} \equiv \sqrt{2\, \alpha^{+}\alpha^{-}} 
= \hat{\gamma}/\sqrt{2 \hat{\gamma}^2 +1} \sim 1/\sqrt{2}$.

When bias $u$ and nonleading parameters 
$(\gamma_{2}, \gamma_{5}, \cdots)$ are turned on,
the zero-modes $\Psi_{0}$ and $\Psi_{1_{\pm}}^{(0)}$
in general deviate from zero energy and become the pseudo-zero-modes.
Their spectra, to first order in such perturbations,  can also be determined 
using this $u\rightarrow0$ zero-mode basis  
$\Psi^{\rm pz} = (\Psi_{0}, \Psi_{1_{+}}^{(0)},  \Psi_{1_{-}}^{(0)})^{\rm t}$.
Writing $H^{\rm tri}$ in the $3\times 3$ matrix form 
${\cal H}^{\rm pz}_{ij} \sim (\Psi^{\rm pz})^{\dag}_{i} {\cal H}_{K} (\Psi^{\rm pz})_{j}$ 
yields the spectrum of the pseudo-zero-mode (PZM) sector,
\begin{eqnarray}
{\cal H}^{\rm pz} &=&\{0\} \oplus {\cal H}_{1}, \nonumber\\
{\cal H}_{1}&=& \sigma u\, \sigma_{3} + \beta_{0}\, 1 + \beta\, \sigma_{1},
\end{eqnarray}
where
\begin{eqnarray}
\beta &=& \textstyle{1\over{2}} (1-c_{1}^2)\, \gamma_{5}  - c_{1}^2\, \Delta'
+ 2\,\sigma\, c_{1} \lambda\,\omega_{c} ,
\nonumber\\
\beta_{0} &=& -  \textstyle{1\over{2}} c_{1}^2\,  \gamma_{5} + (1-c_{1}^2)\, \Delta' 
+ 2\, \sigma\, c_{1} \lambda\,\omega_{c} . 
\end{eqnarray}
This PZM spectrum ${\cal H}^{\rm pz}$,
in the framework of degenerate perturbation theory,
is correct to order linear 
in $(u,\gamma_{5}, \lambda, \Delta')$, which is sufficient for our present purpose.

Diagonalizing ${\cal H}_{1}$ by a rotation within the $\{1_{\pm} \}$ sector,
\begin{eqnarray}
\Psi_{1_{+}} &=& \cos (\theta/2)\,  \Psi_{1_{+}}^{(0)} 
-  \sin (\theta/2)\,  \Psi_{1_{-}}^{(0)}, 
\nonumber\\
\Psi_{1_{-}} &=& \sin (\theta/2)\, \Psi_{1_{+}}^{(0)} 
+ \cos (\theta/2)\,  \Psi_{1_{-}}^{(0)},
\end{eqnarray} 
yields the eigenspectrum 
\begin{equation}
\epsilon_{1_{\pm}} =\beta_{0} \pm \sqrt{\beta^2 +\sigma^2 u^2}
=\beta_{0} \pm |\beta|/\sin \theta,
\label{Eonepm}
\end{equation}
with $\sin \theta =1/\sqrt{1+ \sigma^2u^2/\beta^2 }$ 
and $\cot \theta = - \sigma u/\beta$;
note that $\beta \approx -11.4\, {\rm  meV} < 0$ and
 $\beta_{0} \approx 18.6\, {\rm meV} >0$ 
 for the set~(\ref{Triparameter}) of parameters and at  $B=$20 T.
In particular, for $u\rightarrow +0$ $(\theta\rightarrow\pi/2$)
the spectrum reads
\begin{eqnarray}
\epsilon_{1_{+}} &\stackrel{u\rightarrow0}{=}& 
\beta_{0} + |\beta| 
=\Delta' - \textstyle{1\over{2}} \gamma_{5} \ 
(\sim 30\, {\rm meV}) , \nonumber\\
\epsilon_{1_{- }} &\stackrel{u\rightarrow0}{=}& 
(1-2c_{1}^2)\, ( \textstyle{1\over{2}} \gamma_{5} \! + \Delta')
+ 4 \sigma c_{1} \lambda\,\omega_{c} , 
\end{eqnarray} 
which, for $\hat{\gamma} \rightarrow \infty$, 
recovers an earlier result,~\cite{KMc83}
with 
$c_{1}^2 \rightarrow 1/2$, $\sigma \rightarrow 0$ 
and $\epsilon_{1_{- }}\rightarrow 0$.

Here we wish to discuss possible effects of 
the interlayer coupling
$\gamma_{3} \equiv \gamma_{A_{1} B_{2}} \propto v_{3}$.
It induces transitions 
that go outside the PZM sector, 
as one can verify using the solutions $(\Psi_{0}, \Psi_{1_{\pm}})$.
Accordingly, its contributions to the spectra 
$(\epsilon_{0},\epsilon_{1_{\pm}})$
are only of second order in $v_{3}/v$ and are negligible 
in high magnetic fields.

The Hamiltonian ${\cal H}_{K'}$ at another valley is given by ${\cal H}_{K}$ 
with replacement $\Pi \leftrightarrow -\Pi^{\dag}$.
As for its spectrum one readily finds the following:
The associated eigenmodes $\Psi_{n}^{K'}$ take the form of $\Psi_{n}$ in Eq.~(\ref{Psi_n}), 
with replacement
$|n \rangle \rightarrow |n-2 \rangle$ for $d_{n}^{(2)}$
and $|n-2 \rangle \rightarrow |n \rangle$ for $(b_{n}^{(1)}, b_{n}^{(3)} )$.
The reduced Hamiltonian ${\cal H}_{n}|^{K'}$ is obtained from ${\cal H}_{n}$
in Eq.~(\ref{Hn}) by replacing each $r_{n-1}$ by $- r_{n}$
and each  $r_{n}$ by  $- r_{n-1}$.
One, of course, has to calculate the eigenvectors 
${\bf v}_{n}|^{K'}=(b_{n}^{(1)}, d_{n}^{(1)},...)^{\rm t}|^{K'}$ anew.

Unlike ${\cal H}_{n}$, 
${\cal H}_{n}^{K'}$ has rank 2 for $n=0$ and rank 5 for $n=1$.
This already signals that the PZM spectra
significantly differ between the two valleys. 
For $n=0$ one considers the $2\times 2$ matrix Hamiltonian 
$\hat{\cal H}_{\rm red}|^{K'}  \sim -u\, \sigma_{3} 
+ {1\over{2}}\gamma_{2} \sigma_{1}$,
with eigenmodes  (denoted as $n=0_{\pm})$,
\begin{eqnarray}
\Psi_{0_{+}} &=& 
\big(-\sin (\phi/2)\, |0\rangle, 0, 0,0, \cos (\phi/2)\,  |0\rangle, 0 \big)^{\rm t},
\nonumber\\
\Psi_{0_{-}} &=&  
\big(\cos (\phi/2)\, |0\rangle, 0, 0,0, \sin (\phi/2)\,  |0\rangle, 0 \big)^{\rm t},
\end{eqnarray}
and the associated spectra
\begin{equation}
\epsilon_{0_{\pm}} =\pm \sqrt{ (\gamma_{2}/2)^2 +u^2}
=\pm \textstyle{1\over{2}} |\gamma_{2}|/\sin \phi ,
\end{equation}
where $\sin \phi =1/\sqrt{1 + (2u/\gamma_{2})^2}$
and $\tan \phi = - {1\over{2}} \gamma_{2}/u$.

For $n=1$, ${\cal H}_{n}^{K'}$ has rank 5.  
Of its five eigenvalues, one belongs to the PZM sector, 
two are monolayer-like
with $\epsilon_{\pm1'} \sim \pm \omega_{c}$ and 
two are bilayer-like with $\epsilon_{\pm1''} \sim \pm \sqrt{2} \gamma_{1}$.
In the $u\rightarrow 0$ basis, the zero-energy mode is given by
\begin{equation}
\Psi_{n= 1}  \stackrel{u\rightarrow0}{=} 
c_{1}\,  ( |1\rangle,0, \kappa\,   |0\rangle, 0,  |1\rangle, 0)^{\rm t},
\label{Psinone}
\end{equation}
where $\kappa \equiv 1/\hat{\gamma}$ and 
 $c_{1}\! \equiv  \hat{\gamma}/\sqrt{2 \hat{\gamma}^2\! +1}= 1/\/\sqrt{2 +\kappa^2}$.
Evaluating the expectation value 
$\epsilon_{1} =\Psi_{1}^{\dag} (\omega_{c}\, {\cal H}_{n=1}|^{K'}) \Psi_{1}$
yields the spectrum of the $n=1$ mode, 
\begin{equation}
\epsilon_{1} = 
c_{1}^{2}\, (\gamma_{2} +4 \kappa \lambda\,  \omega_{c} +\kappa^2 \Delta' ),
\end{equation}
correct to order linear in $(u, \gamma_{2}, \gamma_{5}, v_{4}, \Delta')$ as well.

The LLL, i.e, the PZM sector, 
consists of $n \in (0,1_{\pm})$ at valley $K$ 
and of $n\in (0_{\pm},1)$ at valley $K'$; 
there are thus twelve PZM levels differing in spin, valley and orbital.
It is interesting to look into their structure.
For zero bias 
$u\rightarrow +0$ (i.e., $\theta=\phi \rightarrow \pi/2)$, 
$\Psi_{0}$ and $\Psi_{1_{-}}$ at valley $K$ are predominantly 
composed of the orbital mode $|0\rangle$ and $|1\rangle$, respectively, 
residing on the $B$ sites of the middle layer; 
let us denote this feature as 
$\Psi_{0}|^{K} \sim |0\rangle$ on $B_{2}$ and 
$\Psi_{1_{-}}|^{K}  \sim |1\rangle$ on $B_{2}$.
One can further write 
$\Psi_{1_{+}}|^{K}  \sim |0\rangle$ on $B_{1,3}$,
$\Psi_{1}|^{K'}  \sim |1\rangle$ on $A_{1,3}$,
$\Psi_{0_{-}}|^{K'} \sim |0\rangle$ on $A_{1}$, and 
$\Psi_{0_{+}}|^{K'} \sim |0\rangle$ on $A_{3}$.
This naturally explains 
why $\epsilon_{0}$ and $\epsilon_{1_{-}}$ are insensitive 
to  the outer-layer coupling $(\gamma_{2}, \gamma_{5})$ 
and are less sizable.  
In this way, in $ABA$ trilayers the PZM levels 
show valley asymmetry in composition.

\begin{figure}[htbp]
\includegraphics[scale=.55]{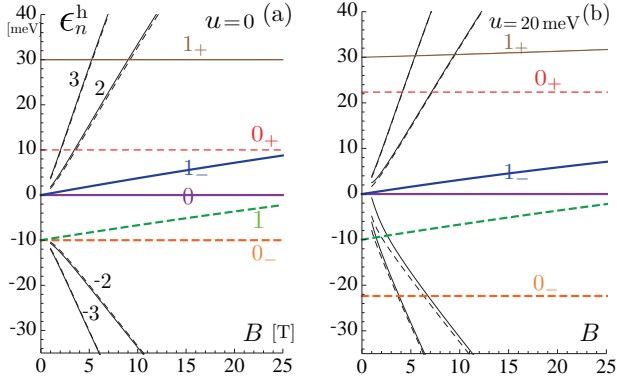}
\caption{ (Color online)
One-body level spectra $\epsilon^{\rm h}_{n}$ 
as a function of $B$ 
for  (a)~$u=0$ and (b)~$u=20$ meV.
Real curves refer to valley $K$ and dashed ones to valley $K'$. 
}
\end{figure}

The valley asymmetry is manifest in the one-body spectra $\{\epsilon_{n}\}$,
which, from now on, are denoted as 
$\{\epsilon_{n}^{\rm h}\}$ to indicate that they come from $H^{\rm tri}$.
Numerically, for $B = 20$\,T and 
with the set of parameters in  Eq.~(\ref{Triparameter}) taken,
\begin{eqnarray}
&&( \epsilon_{1_{+}}^{\rm h}, \epsilon_{1_{-}}^{\rm h}, \epsilon_{0}^{\rm h} )
\stackrel{u\rightarrow0}{\approx} (30, 7.15, 0)\,  {\rm meV}, \nonumber\\
&&(\epsilon_{0_{+}}^{\rm h}, \epsilon_{1}^{\rm h}, \epsilon_{0_{-}}^{\rm h} )
\stackrel{u\rightarrow0}{\approx} (10, -3.64, -10)\,  {\rm meV}.
\end{eqnarray}
Figure~1 shows the spectra $\{ \epsilon_{n}^{\rm h} \}$ 
for $u=(0, 20)$~meV 
as a function of magnetic field $B$, 
along with the $n=\pm 2, \pm 3$ bilayer-like spectra.
The PZM sector is considerably spread in energy
$\sim \Delta' - {1\over{2}}\, ( \gamma_{5}+\gamma_{2}) \sim 40$\,meV,
but, for large $B >15\,$T, it is practically isolated 
from other levels.
The PZM spectra prominently differ 
between the two valleys (real curves {\it vs} dashed ones). 
In addition, they are highly electron-hole ($eh$) asymmetric
(i.e., not symmetric about zero energy) and 
this  $eh$ asymmetry  comes from $\beta_{0} \not =0$ and 
$\epsilon_{1} \not =0$, i.e., primarily from $(\gamma_{5}, \Delta')$ at valley $K$ 
and $\gamma_{2}$ at valley $K'$.
Note that $\epsilon_{1_{\pm}}^{\rm h} $ and $\epsilon_{0_{\pm}}^{\rm h} $ 
vary with interlayer bias $u$. 
Practically only $\epsilon_{0_{\pm}}^{\rm h} $ vary sensitively 
 with $u$ while other levels are barely affected.

Let us now make 
the Landau-level structure explicit 
by passing to
the $|n,y_{0}\rangle$ basis $\ni \{ \Psi_{n} \}$
with $y_{0}\equiv \ell^{2}p_{x}$
via the expansion
$(\Psi^{K} ({\bf x}), \Psi^{K'} ({\bf x}) ) 
= \sum_{n, y_{0}} \langle {\bf x}| n, y_{0}\rangle\, \{ \psi^{n;a}_{\alpha}(y_{0})\}$, 
where $n$ refers to the level index, $\alpha \in (\uparrow, \downarrow)$ 
to the spin, and $a \in (K,K')$ to the valley.
The charge density 
$\rho_{-{\bf p}} =\int d^{2}{\bf x}\,  e^{i {\bf p\cdot x}}\,\rho$ 
with $\rho = (\Psi^{K})^{\dag}\Psi^{K} +  (\Psi^{K'})^{\dag}\Psi^{K'}$ 
is thereby written as~\cite{KSLs}
\begin{eqnarray}
\rho_{-{\bf p}} &=& \gamma_{\bf p}\sum_{k, n =-\infty}^{\infty}
\sum_{a,\alpha} g^{k n;a}_{\bf p}\, 
R^{k n;aa}_{\alpha\alpha;\bf p}, \nonumber\\
R^{kn;ab}_{\alpha\beta;{\bf p}}&\equiv& \int dy_{0}\,
{\psi^{k,a}_{\alpha}}^{\dag}(y_{0})\, e^{i{\bf p\cdot r}}\,
\psi^{n,b}_{\beta} (y_{0}),
\label{chargeoperator}
\end{eqnarray}
where $\gamma_{\bf p} \equiv  e^{- \ell^{2} {\bf p}^{2}/4}$; 
${\bf r} = (i\ell^{2}\partial/\partial y_{0}, y_{0})$
denotes the center coordinate with 
$[r_{x}, r_{y}] =i\ell^{2}$.

The coefficient matrix $g^{k n;a}_{\bf p} \equiv g^{kn}_{\bf p}|^{a}$ 
at valley $a\in (K,K')$ is constructed from the eigenvectors ${\bf v}_{n}|^{a}$,
\begin{eqnarray}
g^{kn}_{\bf p}|^{K} &=&\{ b_{k}^{(1)} b_{n}^{(1)} 
+ b_{k}^{(3)} b_{n}^{(3)} \}\, f_{\bf p}^{|k|-2,|n|-2} 
\nonumber\\
&&+ \{ d_{k}^{(1)} d_{n}^{(1)}  + b_{k}^{(2)} b_{n}^{(2)}  
+ d_{k}^{(3)} d_{n}^{(3)} \}\,  f_{\bf p}^{|k|-1,|n|-1} 
\nonumber\\
&& + d_{k}^{(2)} d_{n}^{(2)}\, f_{\bf p}^{|k|,|n|}, {\rm etc.},
\label{fknp}
\end{eqnarray}
and has the property $(g^{mn;a}_{\bf p})^{\dag}= g^{n,m;a}_{\bf -p}$.
Here
\begin{equation}
f^{k n}_{\bf p} 
= \sqrt{n!/k!}\, (-\bar{q}/\sqrt{2})^{k-n}\, L^{(k-n)}_{n} (|\bar{q}|^{2}/2)
\end{equation}
for $k \ge n \ge 0$, and $f^{n k}_{\bf p} = (f^{k n}_{\bf -p})^{\dag}$;
$\bar{q} =\ell (p_{x}\! -i\, p_{y})$; it is understood that 
$f^{kn}_{\bf p}=0$ for $k<0$ or $n<0$.

Within the PZM sector, 
\begin{eqnarray}
&&g^{00}_{\bf p}=1,\  
g^{01_{\pm}}_{\bf p} 
= (\cos \textstyle{\theta\over{2}} \mp \sin {\theta\over{2}} )\, 
c_{1} \ell\,  p/\sqrt{2}, \nonumber\\
&&g^{1_{\pm}1_{\pm}}_{\bf p} 
= 1  - (1\mp  \sin \theta ) \textstyle{1\over{2}} (c_{1})^2\, \ell^2{\bf p}^2,
\nonumber\\
&&g^{1_{+}1_{-}}_{\bf p} = g^{1_{-}1_{+}}_{\bf p} 
= - (\cos \theta)\,  \textstyle{1\over{2}} (c_{1})^2\, \ell^2 {\bf p}^2,
\label{gpzmK}
\end{eqnarray}
at valley $K$, and 
\begin{eqnarray}
&&g^{0_{+}0_{+}}_{\bf p}=g^{0_{-}0_{-}}_{\bf p} =1,\ \  
g^{0_{+}0_{-}}_{\bf p}=0,  \ \ \nonumber\\
&& g^{0_{\pm}1}_{\bf p} = 
(\cos \textstyle{\phi\over{2}}  \mp \sin {\phi\over{2}} )\, c_{1} \ell\,  p/\sqrt{2},     \ \    \nonumber\\
&& g^{11}_{\bf p} = 1 - (c_{1})^2\, \ell^2 {\bf p}^2,  \ \ \ \ 
\label{gpzmKp}
\end{eqnarray}
at valley $K'$, with $c_{1}= 1/\sqrt{2+ \kappa^2}$
and $\kappa \equiv 1/\hat{\gamma}$.
One can further show that, with only $(v,\gamma_{1}, u)$ kept, 
these $g^{kn}_{\bf p}$ are only corrected to $O(\hat{u}^2\kappa^2)$ or smaller.

In view of Eqs.~(\ref{vn}) and (\ref{Hequi}), 
the form factors $g^{mn;a}_{\bf p}$ enjoy the property
\begin{equation}
g^{mn;a}_{\bf p} = 
g^{-m,-n;a}_{\bf p}|_{-U_{i}, -v_{4}, -\gamma_{2}, -\gamma_{5},-\Delta'}
\label{gmnproperty}
\end{equation}
for general $(m,n)$, where it is understood that 
one sets $\pm m \rightarrow j$ for the PZM level  $j$. 
This property plays a key role in our analysis later.
Equations~(\ref{gpzmK}) and~(\ref{gpzmKp})  
are expressions valid to zero$th$ order in perturbations 
$(u,\gamma_{2}, \gamma_{5},\cdots)$,
but they actually know the nature of perturbations through the mixing angles 
$(\theta, \phi)$ that depend on the relative strengths $(u/\beta, u/\gamma_{2})$.
They indeed satisfy Eq.~(\ref{gmnproperty}).

The form factors $g^{kn}_{\bf p}$ generally differ 
between the two valleys. Interestingly, they happen to coincide 
for $u\rightarrow 0$:
Indeed, for $u\rightarrow 0$, one finds
\begin{eqnarray}
&&g^{1_{+}1_{+}}_{\bf p}=g^{00}_{\bf p}=1,
g^{1_{-}1_{-}}_{\bf p}= 1- c_{1}^2\ell^2 {\bf p}^2, \nonumber\\
&&g^{01_{+}}_{\bf p}=g^{1_{-}1_{+}}_{\bf p}=0,
g^{01_{-}}_{\bf p}= c_{1}\, \ell\, p,
\label{gknU0}
\end{eqnarray}
at valley $K$, and analogous $K'$-valley expressions
with $(1_{+},1_{-},0)$ replaced by $(0_{+},1,0_{-})$ in the above.
This fact tells us that the charge $\rho_{\bf p}$ takes 
a manifestly valley-symmetric form
for zero bias $u=0$ while the one-body spectra 
$\{\epsilon_{n}^{\rm h} \}$ inevitably break valley symmetry.
In addition, Eq.~(\ref{gknU0}) implies that, for $u\rightarrow 0$, 
$1_{+}|^{K}$ is isolated from $(0,1_{-})|^{K}$, and similarly,  
$0_{+}|^{K'}$ from $(1,0_{-})|^{K'}$.

From now on we frequently suppress
summations over levels $n$, spins $\alpha$ and valleys $a$, 
with the convention that the sum is taken over repeated indices.
The Hamiltonian $H^{\rm tri}$ projected to 
the PZM sector is thereby written as
\begin{equation}
H^{\rm h} = \epsilon^{\rm h}_{n}\, R^{nn}_{\alpha\alpha;{\bf p= 0}} 
- \mu_{\rm Z}\, (T_{3})_{\beta\alpha} R^{nn}_{\alpha\beta;{\bf p= 0}} 
\label{Hzero}
\end{equation}
with $n\in (0,1_{\pm}, 0_{\pm}, 1)$. 
Here the Zeeman term $\mu_{\rm Z} \equiv g^{*}\mu_{\rm B}B
\approx 0.12\, B[{\rm T}]$ meV is introduced 
via the spin matrix 
$T_{3} = \sigma_{3}/2$.
Actually, the Zeeman energy $\mu_{\rm Z}$ is only about 3\,meV even at $B=30$\,T
and is generally smaller than energy splitting due to valley breaking.
Accordingly, in what follows, 
we mostly suppose that the spin is practically unresolved 
and focus on energy gaps due to valley and orbital breaking.

\section{vacuum fluctuations}

In this section we examine the effect of Coulombic quantum fluctuations 
on the PZM multiplet.
The Coulomb interaction is written as
\begin{equation}
V
= \textstyle{1\over{2}} \sum_{\bf p}
v_{\bf p}\, :\rho_{\bf -p}\, \rho_{\bf p}:,
\label{Hcoul}
\end{equation}
where
$v_{\bf p}= 2\pi \alpha/(\epsilon_{\rm b} |{\bf p}|)$ 
with 
$\alpha = e^{2}/(4 \pi \epsilon_{0}) \approx 1/137$ and 
the substrate dielectric constant $\epsilon_{\rm b}$.
For simplicity we ignore the difference 
between the intralayer and interlayer Coulomb potentials.

In this paper we generally study many-body ground states $|G\rangle$ 
with a homogeneous density, realized at integer filling factor $\nu \in [-6, 6]$.
We set the expectation values  
$\langle G| R^{mn;ab}_{\alpha\beta; {\bf k}}|G \rangle 
= \delta_{\bf k,0}\, \rho_{0}\,  \nu^{mn;ab}_{\alpha\beta}$
with $\rho_{0} = 1/(2\pi \ell^{2})$, 
so that the filling factor $\nu^{nn;aa}_{\alpha\alpha}=1$ 
for a filled $(n,a,\alpha)$ level.

Let us define the Dirac sea $|{\rm DS}\rangle$ as the valence band 
with levels below the PZM sector 
(i.e., levels with $n\le -2$, $n'\le -1'$, ...) all filled.
We construct the Hartree-Fock Hamiltonian
$V^{\rm HF}= V_{\rm D}+ V_{\rm X}$ out of $V$ as  the effective Hamiltonian 
that governs the electron states over $|{\rm DS}\rangle$.
As usual, the direct interaction $V_{\rm D}$
is removed if one takes into account the neutralizing positive background.  
We thus focus on the exchange interaction
\begin{equation}
V_{\rm X}
= - \sum_{\bf p}v_{\bf p}\gamma_{\bf p}^{2}\,
g^{m n';b}_{\bf -p}\,g^{m'n;a}_{\bf p}\,
\nu^{mn;ba}_{\beta \alpha}\, R^{m' n';ab}_{\alpha\beta;{\bf 0}},
\label{Vex}
\end{equation}
where we sum over filled levels $(m,n)$ 
and retain the PZM sector $m',n'\in (0,1_{\pm}, 0_{\pm}, 1)$.

Let us first extract 
the contribution from the Dirac sea,
\begin{equation}
V^{\rm DS}_{\rm X}
= - \sum_{\bf p}v_{\bf p}\gamma_{\bf p}^{2}\,
\sum_{n\in {\rm DS}}|g^{m'n;a}_{\bf p}|^2\, 
R^{m' m';aa}_{\alpha\alpha;{\bf 0}},
\label{VxDS}
\end{equation}
where the sum is understood over spin $\alpha$ 
and  over $m'\in (0,1_{\pm})$ for $a=K$
and $m'\in (0_{\pm},1)$ for $a=K'$. 
Actually, the sum over infinitely many filled levels 
with $n \in {\rm DS}$
gives rise to  an ultraviolet divergence.

Fortunately this infinite sum is evaluated exactly
to zero$th$ order in perturbations 
$(u,\gamma_{2}, \gamma_{5},\cdots)$,
as done earlier,~\cite{KSLs, KSLsTL}
if one notes Eq.~(\ref{gmnproperty}) and 
the completeness relation~\cite{KSLs} 
\begin{equation}
\sum_{n=-\infty}^{\infty} |g^{mn;a}_{\bf p}|^{2} 
= e^{\ell^{2}{\bf p }^{2}/2}\ \ 
{\rm for\ each}\ a \in (K,K').
 \label{compRel}
\end{equation}
The result is 
\begin{equation}
\sum_{n \in {\rm DS}} |g^{jn;K}_{\bf p}|^{2} 
= {1\over{2}}\, (e^{\ell^{2}{\bf p }^{2}/2} - |g^{j0}_{\bf p}|^{2}
- |g^{j1_{+}}_{\bf p}|^{2}- |g^{j1_{-}}_{\bf p}|^{2}),
\label{DSsum}
\end{equation}
for $j\in (0,1_{\pm})$; analogously for valley $K'$.
The $e^{ \ell^{2}{\bf p }^{2}/2}$ term leads to a divergence 
upon integration over ${\bf p}$; 
it, however, shifts all levels $j$ uniformly and is safely omitted.
The regularized Dirac-sea contribution then reads  
\begin{eqnarray}
V^{\rm DS}_{\rm X}\!\!
&\stackrel{0^{th}}{=}& \! \!
\epsilon^{\rm v}_{0}\, R^{00}_{\alpha\alpha;{\bf 0}}
+ \epsilon^{\rm v}_{1_{+}}\, R^{1_{+}1_{+}}_{\alpha\alpha;{\bf 0}}
+ \epsilon^{\rm v}_{1_{-}}\, R^{1_{-}1_{-}}_{\alpha\alpha;{\bf 0}} 
\nonumber\\
&&\!\!\! 
+ \epsilon^{\rm v}_{0_{+}}\, R^{0_{+}0_{+}}_{\alpha\alpha;{\bf 0}}
+ \epsilon^{\rm v}_{0_{-}}\, R^{0_{-}0_{-}}_{\alpha\alpha;{\bf 0}}
+ \epsilon^{\rm v}_{1}\, R^{11}_{\alpha\alpha;{\bf 0}}, 
\nonumber\\
\epsilon^{\rm v}_{j}\!\!\! &=&\!\!\!\! {1\over{2}} 
\sum_{\bf p}v_{\bf p}\gamma_{\bf p}^{2}
\sum_{n \in (0, 1_{\pm})} |g^{jn}_{\bf p}|^{2} \ {\rm for}\ j \in (0, 1_{\pm});\ \ \ 
\label{VxDStwo}
\end{eqnarray}
analogously for $j \in (0_{\pm},1)$.
Integration over ${\bf p}$, with the formula
$
\sum_{\bf p}v_{\bf p}\gamma_{\bf p}^2 \,[1, (\ell |{\bf p}|)^2, (\ell |{\bf p}|)^4] 
= [1,1,3]\, \tilde{V}_{c} 
$,
yields
\begin{eqnarray}
\epsilon^{\rm v}_{0} &=& 
\textstyle{1\over{2}}\, ( 1 +  c_{1}^2)\, \tilde{V}_{c},
\nonumber\\
\epsilon^{\rm v}_{1_{\pm}}\!\!\! &=& \textstyle{1\over{2}}
 \big[1 - (1\mp \sin \theta)\, {1\over{2}} (c_{1}^2 - 3 c_{1}^4)  
\big]\, \tilde{V}_{c},
\nonumber\\
\epsilon^{\rm v}_{0_{\pm}}\!\!\! &=& 
\textstyle{1\over{2}}\, \big[ 1 + \textstyle{1\over{2}}  c_{1}^2 (1 \mp  \sin \phi) \big]\,
\tilde{V}_{c},
\nonumber\\
\epsilon^{\rm v}_{1} &=& 
\textstyle{1\over{2}}\, (1 - c_{1}^2+ 3\, c_{1}^4 )\, \tilde{V}_{c},
\label{Vzeroreg}
\end{eqnarray}
where  $c_{1}^2\equiv (c_{1})^2$, etc.,  
$\tilde{V}_{c} \equiv \sqrt{\pi/2}\, V_{c}$
and 
\begin{equation}
V_{c} \equiv \alpha/(\epsilon_{b}\ell) \approx (56.1/\epsilon_{b})\, 
\sqrt{B[{\rm T}]}\, {\rm meV}.
\end{equation}
Note that Eqs.~(34) $\sim $ (36) are the zeroth-order expressions, 
which depend on bias $u$ through the zero$th$-order ratios $\sin \phi$ and $\sin \theta$.
Numerically, for $u=0$ and $B = 20$\,T, and 
with $\epsilon_{b}= 5$ taken as a typical value,
\begin{eqnarray}
\epsilon^{\rm v}_{0}&=& \epsilon^{\rm v}_{0_{-}}
=  \textstyle{1\over{2}}\, ( 1 + c_{1}^2)\, \tilde{V}_{c}
\approx 0.73\, \tilde{V}_{c},
\nonumber\\
\epsilon^{\rm v}_{1_{-}}  \!\!\! &=&  \epsilon^{\rm v}_{1} =
\textstyle{1\over{2}}  ( 1 - c_{1}^2+ 3\, c_{1}^4 )\, \tilde{V}_{c}
\approx 0.59\, \tilde{V}_{c}, \nonumber\\
\epsilon^{\rm v}_{1_{+}} \!\!\!
&=& \epsilon^{\rm v}_{0_{+}} = \textstyle{1\over{2}} \, \tilde{V}_{c}
= 0.5\, \tilde{V}_{c}.
\label{Spectrum} 
\end{eqnarray}
In this way, the PZM levels are $\lq\lq$Lamb-shifted"
due to vacuum fluctuations 
and the splitting among $\{\epsilon^{\rm v}_{j}\}$
reflects the difference in their spatial (or ${\bf p}$) distributions.


\begin{figure}[thbp]
\includegraphics[scale=.55]{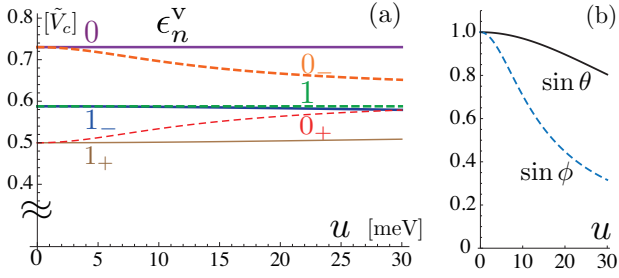}
\caption{ 
(Color online) 
(a) Orbital Lamb-shift corrections  $\{\epsilon^{\rm v}_{j}\}$
plotted in units of $\tilde{V}_{c}$  as a function of bias $u$ at $B=20\,$T.
(b)~$\sin \theta$ and $\sin \phi$ as a function of $u$
for $B=20\,$T.  
}
\end{figure}


In Fig.~2 the Lamb shifts $\{\epsilon^{\rm v}_{j}\}$
are plotted, in units of $\tilde{V}_{c}$,
as a function of bias $u$.
They are ordered, e.g., as $\epsilon_{0}^{\rm v} > \epsilon_{1_{-}}^{\rm v} > \epsilon_{1_{+}}^{\rm v}$
at one valley, and are valley-symmetric for $u=0$
(this reflects the manifest valley symmetry of the charge 
$\rho_{\bf p}$ noted in Sec. II), 
with valley asymmetry developing gradually with increasing bias $u$.

\begin{figure}[tthp]
\includegraphics[scale=.52]{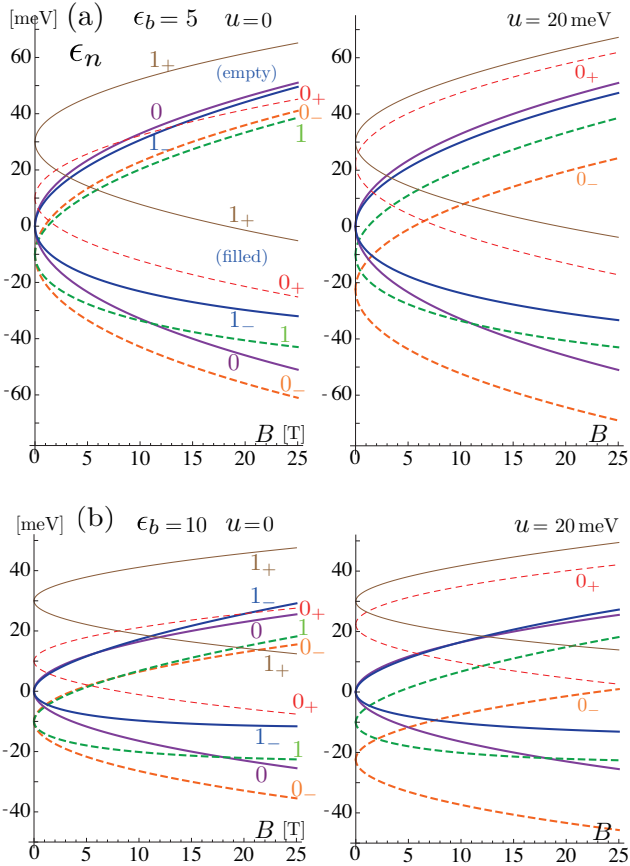}
\caption{ (Color online)  Level spectra of the empty and filled PZM sector.
(a)~$\epsilon_{b}=5$ and $u=(0, 20)$ meV. 
The upper and lower halves of each parabolic spectrum 
refer to the empty ($\nu =-6$) and filled  ($\nu =6$)  case,
respectively.  
(b)~$\epsilon_{b}=10$ and $u=(0, 20)$ meV,
with a weaker Coulomb potential.
}
\end{figure}

The spectra~(\ref{Vzeroreg}) or~(\ref{Spectrum}) refer to 
those of empty levels.
Actually the spectra vary 
(i.e., generally go down due to the exchange interaction) 
with filling of the PZM levels.
In particular,  Eq.~(\ref{VxDS}) tells us that, 
when the PZM sector is filled up, 
$\{\epsilon^{\rm v}_{j}\}$ change sign 
$\epsilon^{\rm v}_{j} \rightarrow - |\epsilon^{\rm v}_{j}|$.
In this sense, the Lamb-shift corrections 
$\{\epsilon^{\rm v}_{j}\}$ preserve $eh$ symmetry.
Thus, as the PZM sector is filled from $\nu=-6$ (empty) to
$\nu=6$ (full), the spectrum of the $j$-th level varies from   
$\epsilon^{\rm h}_{j} + \epsilon^{\rm v}_{j}$ 
to $\epsilon^{\rm h}_{j} - \epsilon^{\rm v}_{j}$. 
See Fig.~3(a),
which depicts the (empty/filled) spectra 
$\epsilon^{\rm h}_{j} \pm \epsilon^{\rm v}_{j}$ 
as a function of $B$, for $u=0$ and $\epsilon_{\rm b} =5$; 
the upper and lower halves of each spectrum refer
to empty and filled levels, respectively. 
Note that bias $u$ works to enhance 
the splitting of the $0_{\pm}$ spectra. 
For reference, Fig.~3(b) shows the level spectra 
for a weaker potential with $\epsilon_{\rm b} =10$.

 At valley $K$, the one-body spectra
$\{\epsilon_{n}^{\rm h} \}$ are ordered so that 
$\epsilon_{1_{+}}^{\rm h} > \epsilon_{1_{-}}^{\rm h} > \epsilon_{0}^{\rm h}$
while  $\{\epsilon_{n}^{\rm v} \}$ are ordered so that 
$\epsilon_{0}^{\rm v} > \epsilon_{1_{-}}^{\rm v} > \epsilon_{1_{+}}^{\rm v}$.
The Lamb-shift contributions $\{\epsilon_{n}^{\rm v} \}$ 
therefore enhance splitting among filled levels  $(1_{+}, 1_{-}, 0)$
with spectra $\epsilon_{n}^{\rm h} - \epsilon_{n}^{\rm v}$.
For empty levels (with $\epsilon_{n}^{\rm h} + \epsilon_{n}^{\rm v}$) 
they work oppositely and even reverse the ordering 
of the $1_{-}$ and $0$ spectra 
when the Coulomb interaction $\tilde{V}_{c}$ is strong enough
(i.e., for smaller $\epsilon_{b}$ and higher $B$); 
compare Figs.~3(a) and 3(b).
Replacing $(1_{+}, 1_{-}, 0) \rightarrow (0_{+}, 1, 0_{-})$
also allows one to find essentially the same features for valley $K'$.
In this way the Lamb-shift contributions, 
though $eh$ symmetric by themselves, 
work to enhance $eh$ asymmetry 
in the full PZM spectra $\epsilon^{\rm h}_{j} \pm \epsilon^{\rm v}_{j}$.

To see how each level evolves with filling, 
one has to examine the Coulomb interaction acting within the LLL;
we study this in the next section.

\section{Coulomb interactions}
 
The Coulomb exchange interaction 
acting within the PZM sector is written as 
\begin{equation}
V_{\rm X}^{\rm pz} =  - \tilde{V}_{c}\, \Gamma^{n' m'}_{\beta\alpha}\, 
R^{m' n'}_{\alpha\beta;{\bf 0}}, 
\end{equation}
with
$\Gamma^{n' m'}_{\beta\alpha} \equiv
\sum_{\bf p} v_{\bf p}\gamma_{\bf p}^{2}\, 
(g^{n' m}_{\bf p})^{*}\,g^{m'n}_{\bf p}\,
\nu^{mn}_{\beta \alpha}/\tilde{V}_{c} = \Gamma^{m' n'}_{\alpha\beta}$,
where $(m', n')$  and $(m,n)$ are summed over  $(0,1_{\pm}, 0_{\pm},1)$.
For definiteness we focus on the $u\rightarrow 0$ case,
where $g^{mn}_{\bf p}$ and $\Gamma^{n' m'}_{\beta\alpha}$ 
considerably simplify.
Indeed, as noted in Eq.~(\ref{gknU0}),
for $u=0$, the charge $\rho_{\bf p}$ takes a valley-symmetric form
and, in addition, one has 
$g^{01_{+}}_{\bf p} =g^{1_{-}1_{+}}_{\bf p} =0$
and $g^{0_{+}0_{-}}_{\bf p}=g^{0_{+}1}_{\bf p} =0$.
This structure suggests that, at valley $K$, 
$1_{+}$ tends to be isolated from 
$(0, 1_{-})$ which may potentially get mixed;
similarly, $0_{+}$ tends to be isolated from $(0_{-},1)$ at valley $K'$.
Actually, for $u=0$ one finds that
\begin{eqnarray}
&&\Gamma^{00} = \nu^{00} + \nu^{1_{-}1_{-}}\, c_{1}^{2},\  
 \nonumber\\
 &&\Gamma^{1_{-}1_{-}} = \nu^{00}\, c_{1}^{2}
 + \nu^{1_{-}1_{-}}\, (1- 2 c_{1}^{2} + 3 c_{1}^{4}), \nonumber\\
&&\Gamma^{01_{-}} = \nu^{01_{-}}\, (1-  c_{1}^{2}),\ 
\Gamma^{1_{+}1_{-}} = \nu^{1_{+}1_{-}}\, (1-  c_{1}^{2}), \nonumber\\
&&\Gamma^{1_{+}1_{+}} = \nu^{1_{+}1_{+}}, \
 \Gamma^{01_{+}} = \nu^{01_{+}},
\label{GatK}
\end{eqnarray}
at valley $K$, with obvious spin indices $(\alpha,\beta)$ suppressed.
Replacing $(0,1_{-},1_{+}) \rightarrow (0_{-},1,0_{+})$ in the above 
yields expressions $\Gamma^{K'K'}$ for valley $K'$ 
and, in an analogous way, 
mixed-valley components $\Gamma^{KK'}$ as well. 
One can further use new orbital labels $\hat{n}$ 
and rename $(0,1_{-},1_{+})$ as $(\hat{1},\hat{2}, \hat{3})$
with valley $a=K$, 
and $(0_{-}, 1, 0_{+})$ as $(\hat{1},\hat{2}, \hat{3})$
with valley $a=K'$, 
so that, e.g., $\nu^{00}=\nu^{\hat{1}\hat{1}; KK}$, 
$\nu^{00_{-}}=\nu^{\hat{1}\hat{1}; KK'}$, etc.
Then the exchange interaction $V_{\rm X}^{\rm pz}$ itself 
is cast into a valley- (and spin-)symmetric form 
composed of terms like $\nu^{\hat{n} \hat{m};ba}_{\beta\alpha}\, 
R^{\hat{m}' \hat{n}';ab}_{\alpha\beta;{\bf 0}}$.

The PZM levels 
are now governed by the effective Hamiltonian
${\cal V} \equiv H^{\rm h} +V^{\rm DS}_{\rm X} + V_{\rm X}^{\rm pz}$.
Note first that the interaction $V^{\rm DS}_{\rm X} + V_{\rm X}^{\rm pz}$
is symmetric in spin and valley (for $u=0$). 
Thus ${\cal V}$ is made diagonal in valley and spin 
if one takes the valley basis $(K, K')$ and 
the spin basis $(\uparrow, \downarrow)$
of the one-body part $H^{\rm h} \sim \{\epsilon_{n}^{\rm h}\}$.
Accordingly, one can treat each valley and spin separately, 
and diagonalize ${\cal V}$ with respect to the orbital modes 
$(0, 1_{-}, 1_{+})|^{K}$ and $(0_{-},1, 0_{+})|^{K'}$ for each spin.

Let us  now discuss how the PZM levels evolve with filling.
For definiteness, we first suppose filling the empty PZM sector 
with electrons gradually 
under a fixed magnetic field $B=20$\,T and $u=0$.
We consider 6 levels, $(0,1_{\pm},0_{\pm}, 1)$ per spin,
and use $0 \le n_{\rm f} \le 6$ to denote the filling factor 
for this subsector; with electron spins supposed to be unresolved,  
the PZM sector thereby has 
the filling factor $\nu = 2 (n_{\rm f}-3)$.
(We refer to the case of resolved spins later.)
Our focus is on uniform ground states at integer filling. 
To follow their evolution, we choose 
to diagonalize the Hamiltonian ${\cal V}$
with uniform states at intermediate filling factors; 
this serves to visualize how level mixing and crossing take place, 
as we shall see.

One can read from Fig.~3(a) the level spectra of the empty/filled PZM sector,
 \begin{eqnarray}
&&(\ \epsilon_{1}, \ \ \epsilon_{0_{-}},\ \ \epsilon_{0_{+}},\  
\epsilon_{1_{-}},\ \ \epsilon_{0},\ \,  \epsilon_{1_{+}} ) 
\nonumber\\
&\stackrel{\rm empty}{\approx}& (33.3, 35.9, 41.4, 44.1, 45.9, 61.4)\ {\rm meV},
\nonumber\\
&\stackrel{\rm filled}{\rightarrow}& \!\!\!\!
- (40.6, 55.9, 21.4, 29.8, 45.9, 1.44)\ {\rm meV},
\label{emptypzmspec}
\end{eqnarray}
where $\epsilon_{n}|^{\rm empty/filled} 
= \epsilon^{\rm h}_{n} \pm \epsilon^{\rm v}_{n}$
and $\epsilon_{b}=5$.
It is seen from these spectra that
the $1|^{K'}$ and $0_{-}|^{K'}$ levels potentially have crossing,
so do $0|^{K}$ and $1_{-}|^{K}$.
 This signals level mixing within each pair.

It is the  lowest-lying $1|^{K'}$ level that starts to be filled first.
As it is filled, it cooperates with the $0_{-}|^{K'}$ level, 
paired via the exchange interaction.
To clarify how they evolve, one can now try to diagonalize 
${\cal V} = H^{\rm h} +V^{\rm DS}_{\rm X} + V_{\rm X}^{\rm pz}$
for the three lowlying levels
$(1, 0_{-}, 0_{+})|^{K'}$, and subsequently 
for those at valley $K$.
See the appendix for an analysis.


\begin{figure}[thbp]
\includegraphics[scale=.7]{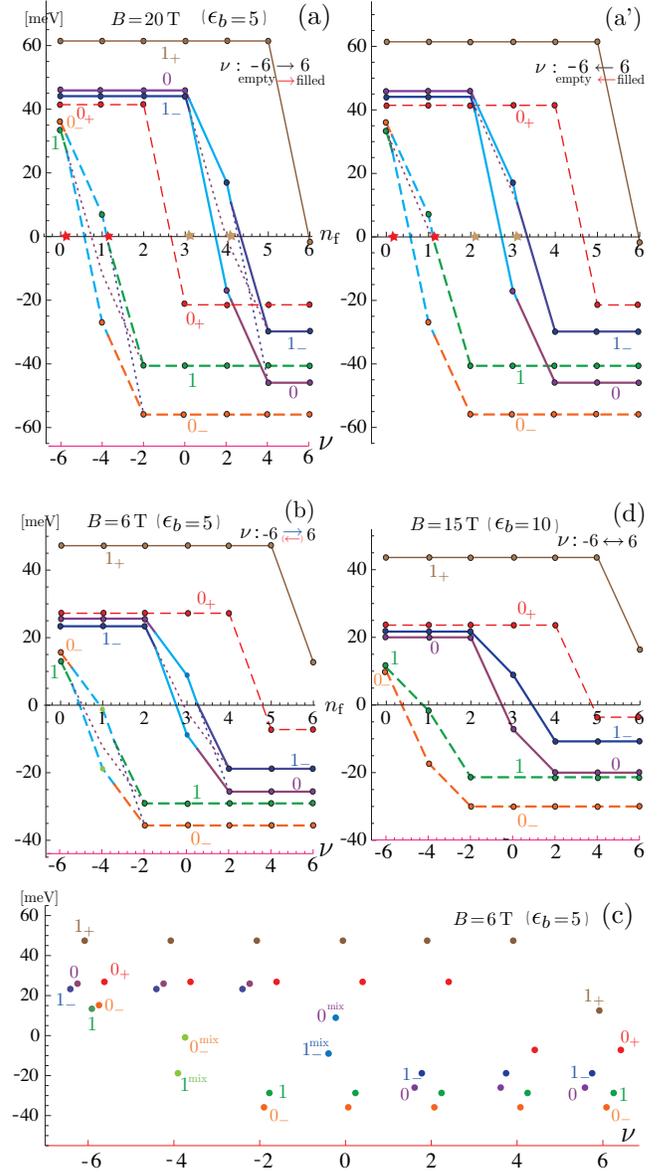}
\caption{(Color online)
(a) Evolution of  level spectra 
at $B=20\,$T and $\epsilon_{b} =5$ as the PZM sector is filled from 
$n_{\rm f}=0$ (empty) to $n_{\rm f}=6$ (full);
the filling factor $\nu=2(n_{\rm f} - 3) \in [-6,6]$, 
with the electron spin supposed to be unresolved.
Orbital mixing takes place over the interval 
$n_{\rm f} \in (0.145, 1.145)$ and $n_{\rm f} \in (3.101, 4.101)$, 
indicated by colored stars. 
Thin dotted curves represent evolution of level spectra
when no orbital rotation were allowed.
(a')~Evolution of level spectra when the PZM sector is emptied from 
$n_{\rm f}=6$ to $n_{\rm f}=0$.
(b) and (c)~Level spectra for $B=6\, {\rm T}$ and  $\epsilon_{b} =5$.
(d)~Level mixing is absent 
for a weaker Coulomb potential with $\epsilon_{b} =10$
and at $B=15\,$T. 
}
\end{figure}


Figure 4(a) summarizes the resulting evolution 
of level spectra.
For $0 \le n_{\rm f} \le  n_{\rm cr}$ with $n_{\rm cr} \approx 0.145$
only the $1|^{K'}$ level is filled and 
gets lower in energy along with the (paired) empty $0_{-}|^{K'}$ level.
Beyond $n_{\rm cr}$, $1|^{K'}$ is mixed with $0_{-}|^{K'}$ 
and turns into $0_{-}|^{K'}$
at $n_{\rm f} \approx 1.145$;
at the same time, $0_{-}|^{K'}$ turns into $1|^{K'}$.
Both $0_{-}|^{K'}$ and $1|^{K'}$  levels 
are eventually filled up at $n_{\rm f}=2$.
At integer filling $n_{\rm f}=1$, 
the $(1,0_{-})$-mixed levels consist of 
a filled level of energy $\approx -27.0$ meV and 
an empty level of energy $\approx 7.0$ meV.
A close look into Fig.~4(a) reveals that 
an $\lq\lq$orbital" rotation takes place 
so as to avoid level crossing. 
The remaining $0_{+}|^{K'}$ level evolves individually,
and is filled over the interval $n_{\rm f} \in [2,3]$.

For $n_{\rm f} >3$ an analogous process is repeated 
for $(1_{-},0,1_{+})|^{K}$ levels at another valley. 
There mixing of $1_{-}|^{K}$ and $0|^{K}$ takes place 
over the interval $n_{\rm f} \in (n'_{\rm cr}, 1+ n'_{\rm cr} )$ 
with $n'_{\rm cr} \approx 3.101$, and avoids level crossing.

From Fig.~4(a) one can read off the spectra of the PZM sector 
at each integer filling factor $\nu \in [-6,6]$.
The spectra are $eh$- and valley-asymmetric.
Let us now note that, due to this $eh$ asymmetry, 
the level spectra may evolve in a different pattern 
when one empties the PZM sector rather than filling it.
Indeed, such a difference is clearly seen from 
Fig.~4(a'), which shows the evolution of level spectra 
when the filled PZM sector is gradually emptied 
(i.e., $\nu= 6 \rightarrow -6$) under the same $B=20$\,T. 
Actually, Fig.~4(a') is a result of direct calculation, 
but it will be clear how to draw it by a glance at Fig.~4(a).

For comparison, see also Figs.~4(b) and 4(c), 
which show the evolution of level spectra under $B=6\,$T.
There the pattern of evolution is uniquely fixed, 
independent of 
whether one fills or empties the PZM sector.
Lastly, Fig.~4(d) illustrates the case of 
a weaker Coulomb potential
with $\epsilon_{\rm b} =10$ and at $B=15\,$T, 
corresponding to the level spectra in Fig.~3(b).  
Here again the pattern of level spectra is 
uniquely fixed, but, unlike in the above cases, 
there is no level mixing.

In general, the level spectra 
$\epsilon^{\rm h}_{j} \pm \epsilon^{\rm v}_{j}$ 
of the empty/filled PZM sector 
(in Fig.~3) are fixed in advance
by specifying the value of magnetic field $B$ 
at $\nu=-6$ and $\nu=6$, respectively.
How the spectra evolve at intermediate filling factors, as we have seen, 
depends on whether one fills or empties the PZM sector
and how one does it, 
e.g., under fixed $B$ or fixed density $\rho \propto \nu B$.

It will be clear from the model calculations above that 
the $1_{+}|^{K}$ and $0_{+}|^{K'}$ levels 
evolve individually without mixing with others
while $1_{-}|^{K}$ and $0|^{K}$ move in pairs, 
so do $(1, 0_{-})|^{K'}$.
Actually, with this experience,
a close look into the empty/filled spectra in Fig.~3 
allows one to draw a general idea about
how the level spectra evolve under fixed $B$ and 
$u=0$ (or even for small $u$ as well).

For example, 
the presence or absence of level mixing is inferred from  Fig.~3.
Level mixing takes place so as to avoid crossing 
of paired levels $(1_{-}, 0)|^{K}$ or $(1,0_{-})|^{K'}$.
As noted in Sec.~II, the ordering of these paired levels, 
i.e., $\epsilon_{1_{-}}^{\rm h} > \epsilon_{0}^{\rm h}$
and $\epsilon_{1}^{\rm h} > \epsilon_{0_{-}}^{\rm h}$ for
$\epsilon_{n}^{\rm h}$, 
is reversed for the full spectra 
$\epsilon_{n}^{\rm h}+\epsilon_{n}^{\rm v}$
when the Coulomb potential 
$\tilde{V}_{c} \sim \alpha/(\epsilon_{b}\ell)$ is strong enough.
It is thus this inversion of (empty) paired levels 
that drives level mixing.
Accordingly, with $\epsilon_{\rm b}=5$, 
mixing of paired levels is necessarily present 
for almost all values of $B$ in Fig.~3(a), 
as indeed seen from Figs.~4(a) and 4(b).
For Fig.~3(b), 
i.e., for a weaker potential with $\epsilon_{\rm b} =10$, 
mixing is present only at low  $B \lesssim 7\, $T and 
is absent at higher $B$,
as is the case with Fig.~4(d).
When bias $u$ is turned on, 
mixing of $(1, 0_{-})|^{K'}$ disappears rapidly
with increasing $u$, 
but mixing of $(1_{-}, 0)|^{K}$ tends to persist at low $B$,
as verified easily.

In the level spectra of Fig.~3, 
the $0_{+}|^{K'}$ level is relatively isolated upward 
from the paired levels $(1,0_{-})|^{K'}$, 
so is $1_{+}|^{K}$ from $(1_{-},0)|^{K}$.
It is therefore the lowest-lying $(1, 0_{-})|^{K'}$ pair 
that is filled first as $n_{\rm f}=0 \rightarrow 2$  
(or emptied last as $n_{\rm f}=2 \rightarrow 0$).
This leads to a unique $\nu=-2$ ground state
[consisting of filled $(1,0_{-})|^{K'}$ levels] 
with a relatively large $\nu=-2$ level gap, 
as is evident from Fig.~4.
Likewise, an isolated $1_{+}|^{K}$ level leads to 
a unique $\nu=4$ state 
with a relatively large gap.
The ground states at other filling factors, in contrast, 
vary in composition case by case. 
In particular, one notices an equally large $\nu=2$ gap
and a relatively small $\nu=0$ gap 
in the spectra of Figs.~4(b) and 4(d), 
which show essentially the same low-$B$ characteristics 
of the PZM sector 
[at $B\lesssim 10 {\rm T}$ in Fig.~3(a) 
or $\lesssim 20{\rm T}$ in Fig.~3(b)].
Interestingly, in those low-$B$ cases, the $\nu=0$ state 
(essentially) consists of  filled $(0,1,0_{-})$ levels, 
which is the same in composition as the $\nu=0$ state  
one naively expects from the one-body spectra 
$\{\epsilon_{n}^{\rm h}\}$ in Fig.~1 alone.

We have so far supposed unresolved electron spins.  
Note that the exchange interaction acts on pairs of the same spin and valley. 
Accordingly, if, e.g., in Fig.~4(b), 
there were two $(1, 0_{-})|^{K'}$ pairs of spin up and down 
resolved against possible disorder, 
each pair would repeat the $n_{\rm f}=0 \rightarrow 2$ evolution in the figure 
over the interval $\nu= - 6 \rightarrow -4 \rightarrow -2$,
yielding a $\nu=-4$ gap comparable to the $\nu=-2$ gap.
In this way, 
small spin gaps, if resolved, are equally well enhanced by the interaction,
and will modify the evolution patterns in Fig.~4 accordingly.

The transport properties  of graphene trilayers have been studied 
in a number of 
experiments,~\cite{BZZL,KEPF,TWTJ,LLMC,BJVL,ZZCK,JCST,HNE,LVTZ}
and some nontrivial features of the LLL 
of $ABA$-stacked trilayers have been observed.
Evidence for the opening of the $\nu=0$ gap 
comes from early observations~\cite{BZZL,BJVL} 
of an insulating $\nu=0$ state 
in both $ABA$ and $ABC$ trilayers.

Recent experiments on substrate-supported $ABA$ trilayer graphene  
by Henriksen {\it et al.}~\cite{HNE} observed 
a robust $\nu=-2$ Hall plateau and 
a possible incipient $\nu=2$ or $\nu=4$ plateau 
under zero bias ($u\sim 0)$, and 
$\nu=\pm2, \pm 4$ plateaus in biased samples.
Subsequent measurements on dual-gated suspended devices  
by Lee {\it et al.}~\cite{LVTZ} 
observed $\nu= \pm 2$ plateaus
at low magnetic field $B < 4\, {\rm T}$ and
also resolved, in high magnetic fields, additional plateaus 
 at $\nu= \pm 1, \pm 3, -4$, and $- 5$, 
 indicating almost complete lifting of
the 12-fold degeneracy of the LLL. 
Common to these observations, in particular, 
is $eh$ asymmetry 
in the sequence of plateaus, with a prominent $\nu=-2$ plateau.

The $B=6\, {\rm T}$ and $\epsilon_{b}=5$ case of Fig.~4(b)
appears to capture these features seen in experiment at low $B$.
For resolved spins, this case will lead to large gaps 
at $\nu= \pm 2, \pm 4$, 0, 3 and 5, and  
relatively small gaps at $\nu=\pm 1$, -3 and -5.
In our picture, appreciable $eh$ asymmetry is 
a result of Coulombic enhancement 
of the $eh$ asymmetry in $H^{\rm h}$ 
and a large $\nu=-2$ gap is triggered by the valley asymmetry
of $H^{\rm h}$ such that 
${\rm min}[\epsilon_{0_{-}}^{\rm h}, \epsilon_{1}^{\rm h}]^{K'} 
< {\rm min}[\epsilon_{0}^{\rm h}, \epsilon_{1_{-}}^{\rm h} ]^{K}$,
i.e., the $K'$  valley is relatively lower in spectrum.
In general, large level gaps are associated 
with evolution of orbital modes 
$(1_{-}, 0)|^{K}$ and $(1, 0_{-})|^{K'}$ of basic filling step 
$\Delta n_{\rm f} =2$ per spin and 
evolution of rather independent modes $1_{+}|^{K}$
and $0_{+}|^{K'}$ of step $\Delta n_{\rm f} =1$.
This is in sharp contrast to the case of $ABC$ trilayers, 
where large gaps within the LLL are associated 
with evolution (actually, mixing) 
of orbital modes of basic step~\cite{KSLsTL} 
$\Delta n_{\rm f} =3$ per spin, 
leading to visible $\nu=0, \pm 3$ plateaus.

\section{Summary and discussion}

In a magnetic field graphene trilayers develop, 
as the LLL, a multiplet of twelve nearly-zero-energy levels 
with a three-fold orbital degeneracy. 
In this paper we have examined the quantum characteristics 
of this PZM multiplet 
in $ABA$ trilayers, with the Coulomb interaction and
the orbital Lamb shift
taken into account.
It turned out that $ABA$ trilayers are distinct 
in zero-mode characteristics from $ABC$ trilayers 
examined earlier.~\cite{KSLsTL}
We have, in particular, seen that 
both valley and $eh$ symmetries 
are markedly broken in the LLL of $ABA$ trilayers.
These asymmetries appear  
in the one-body spectra $\{\epsilon^{\rm h}_{n}\}$ 
already to first order in nonleading hopping parameters 
(such as  $\gamma_{2}$, $\gamma_{5}$ and $\Delta'$),
and are enhanced 
via the Lamb-shift contributions $\{\epsilon^{\rm v}_{n}\}$ 
and the Coulomb interaction acting within the LLL.

In contrast, for $ABC$ trilayers
the one-body PZM spectra $\{\epsilon^{\rm h}_{n}\}$ 
involve, for  zero bias $u=0$, only a tiny $eh$ asymmetry 
of $O(v_{4}) \sim O(v_{4} \omega_{c}/v \tilde{\gamma})$ 
and no valley breaking~\cite{fntwo} linear in 
nonleading parameters $(\gamma_{2}, \gamma_{3},  v_{4})$.
The Lamb-shift corrections $\{\epsilon^{\rm v}_{n}\}$
add no further breaking (to the leading order). 
Accordingly, in $ABC$ trilayers the LLL is far less afflicted
by $eh$ and valley asymmetries.

The PZM levels differ in structure between the two types of trilayers.
In $ABA$ trilayers they are composed of 
the $|0\rangle$ and $|1\rangle$ orbital modes distributed
in a distinct way at each valley,
as noted in Sec.~II, and in this sense 
the associated valley asymmetry is intrinsic. 
In contrast, in $ABC$ trilayers
these levels are characterized 
by the $|0\rangle$, $|1\rangle$ and  $|2\rangle$
orbital modes residing predominantly 
on one of the outer layers,~\cite{KSLsTL}
with the two valleys related symmetrically 
[by layer and site interchange 
$(A_{1}, B_{1}) \leftrightarrow (B_{3}, A_{3})]$.

The two types of trilayers substantially differ in the way 
the Coulomb interaction acts within the LLL.
They thus differ in the way  
large level gaps or 
the associated conductance plateaus 
appear  within the LLL, 
with $ABA$ trilayers having basic filling steps of $\Delta n_{\rm f} =(2,1)$
and $ABC$ trilayers having a step of $\Delta n_{\rm f} =3$.

Interlayer bias $u$ also acts quite differently
on the two types of trilayers.
For $ABC$ trilayers, $u$ acts oppositely at the two valleys 
and enhances valley gaps.
In contrast, for $ABA$ trilayers, it works to further split  
$\epsilon_{0_{\pm}}|^{K'}$
(and $\epsilon_{1_{\pm}}|^{K}$), i.e., 
enhance orbital breaking at each valley.

The orbital Lamb shift is a many-body vacuum effect 
but is intimately correlated with the Coulomb interaction 
acting within the multiplet.
This is clear if one notes that the filled PZM sector and the empty one, 
both subject to quantum fluctuations of the filled valence band, 
differ by the amount of this Coulomb interaction.
It will be clear now why this vacuum effect, 
though it could easily be overlooked if one naively relies 
on the Coulomb interaction projected to the LLL alone,
has to be properly taken into account in every attempt 
to explore the PZM sector in graphene few-layers.~\cite{Toke}

The $eh$ and valley asymmetries inherent to $ABA$ trilayers
substantially modify the electron and hole spectra within the LLL. 
The sequence of broken-symmetry states, 
observable via the quantum Hall effect, 
is thereby both $eh$- and valley-asymmetric and can change in pattern, 
depending on how one fills or empties the LLL.
We have presented some model calculations in Sec.~IV, 
assuming a typical set~(\ref{Triparameter}) of parameters
and $\epsilon_{b}$.
They are intended to illustrate 
what would generally happen
when the orbital Lamb shift and Coulomb interactions 
are properly taken into account.
They will also be a good base point 
for a more elaborate analysis 
when more data on graphene trilayers 
become available via future experiments.

\acknowledgments

This work was supported in part 
by a Grant-in-Aid for Scientific Research
from the Ministry of Education, Science, Sports and Culture of Japan 
(Grant No. 24540270).

\appendix

\section{Evolution of level spectra}

In this appendix 
we diagonalize the effective hamiltonian 
${\cal V}\equiv H^{\rm h} +V^{\rm DS}_{\rm X} + V_{\rm X}^{\rm pz}$
for the three lowlying levels
$(0_{+}, 0_{-},1)|^{K'}$,
with the electron spin kept frozen.
Let us write 
${\cal V} = H^{mn} R^{mn}_{\bf p=0}$
and
$H^{mn} \equiv (\epsilon_{n}^{\rm h} + \epsilon_{n}^{\rm v} )\,  \delta^{mn} 
- \tilde{V}_{c}\,\Gamma^{nm}$, 
with $m, n \in (1,2,3)$ for $(0_{+},0_{-},1)$.
Note that $\Gamma^{mn}$ are real for real filling factors $\nu^{mn}$, 
which we take.
It therefore suffices to use a real $O(3)$ rotation
to diagonalize the $3 \times 3$ real symmetric matrix  $H^{mn}$.
We thus rotate 
$\psi^{m}$ in orbital $(0_{+},0_{-},1)$ space,
$\psi^{m}= {\cal U}^{m n}\, \phi^{n}$,
with  three Euler angles $(\theta_{3},\theta_{2},\theta_{1})$ 
parameterizing
\begin{equation}
{\cal U}
= e^{-i \theta_{3}\, t_{3}} e^{-i \theta_{2}\, t_{2}}  e^{-i \theta_{1}\, t_{1}},
\end{equation}
where the spin-1 generators $(t_{a})^{bc} \equiv i \epsilon^{bac}$ 
in terms of the totally antisymmetric tensor 
$\epsilon^{abc}$ with $ \epsilon^{123} =1$.
Note that $\theta_{1}$ mixes $n=(2,3)$, $\theta_{2}$ mixes $(1,3)$, etc.
We make ${\cal H} = {\cal U}^{\dag} H\,  {\cal U}$
diagonal,
using the filling factors $\nu^{mn} =({\cal U}^{mn'} )^*  N_{n'}  {\cal U}^{nn'}$
expressed in terms of the filling fraction $N_{n} =(N_{1},N_{2},N_{3})$ 
of the diagonalized levels $\phi^{n}$.

Let us start filling the empty PZM sector at 
(relative) filling factor $n_{\rm f}=0$.
Obviously, 
it is the lowest-lying $1|^{K'}$ level 
that starts to be filled.
To follow how it evolves let us suppose 
that it is filled with fraction  $n_{\rm f}\le 1$ 
and set $(N_{1}, N_{2}, N_{3}) = (0,0,n_{\rm f})$.
${\cal H}^{mn}$ is diagonalized if one can adjust 
$\{ \theta_{n} \}$ so that 
${\cal H}^{12} = {\cal H}^{13}= {\cal H}^{23}=0$.

The onset of possible rotations is seen from the behavior of 
these $({\cal H}^{12} ,  {\cal H}^{13},  {\cal H}^{23})$
under small rotations.
 To first order in $\{ \theta_{n} \}$, 
\begin{eqnarray}
{\cal H}^{12} &\approx& 
-(5.526 + 28.95\, n_{\rm f})\,  \theta_{3} + \cdots,\nonumber\\ 
{\cal H}^{13} &\approx&  
(8.124 + 11.03\, n_{\rm f})\,  \theta_{2}  + \cdots, \nonumber\\ 
{\cal H}^{23} &\approx&  (-2.597 + 17.91\,n_{\rm f})\, \theta_{1}  + \cdots.
\end{eqnarray}
This structure reveals that 
$\{\theta_{n} \}=0$ for $0 \le n_{\rm f}< n_{\rm cr}$
with $n_{\rm cr} \approx 2.597/17.91 \approx 0.145$ 
while $\theta_{1}\not=0$ is possible for $n_{\rm f} > n_{\rm cr}$.
Solving for   $\{ \theta_{n} \}$ numerically
shows that the energy eigenvalue ${\cal H}^{33}$ 
is indeed lowered for $n_{\rm f} > n_{\rm cr}$
with $\theta_{1} \not=0$ and $\theta_{2}= \theta_{3}=0$.
A further analysis reveals that $\theta_{1}$ rises 
from 0 to $\pi/2$ for 
$n_{\rm c} \le n_{\rm f} \le n_{\rm c}^{+} \approx 1.145$ 
and then keeps $\pi/2$ up to $n_{\rm f}=2$.  
(For $1 \le n_{\rm f} \le 2$ we set  $(N_{1}, N_{2}, N_{3}) = (0,n_{\rm f}-1,1)$ 
and extend the solution across $n_{\rm f} = 1$.)
Thus, over the interval $n_{\rm f} \in (n_{\rm cr}, n_{\rm cr}^{+})$
a rotation takes place in orbital space, and  thereby
the $1|^{K'}$ and $0_{-}|^{K'}$ levels are interchanged. 
Finally, the remaining $0_{+}$ level is filled individually for $2< n_{\rm f} \le 3$.

A similar analysis is also made 
for the $(1_{+},0,1_{-})|^{K}$ levels at another valley.
The resulting evolution of level spectra is summarized in Fig.~4(a).



\begin{thebibliography}{99}

\bibitem{MF} E. McCann and V. I. Fal'ko, Phys. Rev. Lett. {\bf 96}, 086805 (2006).


\bibitem{OBSHR} T. Ohta, A. Bostwick, T. Seyller, K. Horn, and E. Rotenberg,
Science {\bf 313}, 951 (2006).


\bibitem{Mc} E. McCann, Phys. Rev. B {\bf 74},  161403(R) (2006).


\bibitem{CNMPL} E.~V. Castro, K.~S. Novoselov, S.~V. Morozov, 
N.~M.~R. Peres, J.~M.~B.~Lopes dos Santos, J. Nilsson, 
F. Guinea, A.~K. Geim, and A.~H. Castro Neto, 
Phys. Rev. Lett. {\bf 99}, 216802 (2007). 


\bibitem{velrenorm} J. Gonz\'{a}lez, F. Guinea, and M.~A.~H. Vozmediano, 
Nucl. Phys. B {\bf 424}, 595 (1994).


\bibitem{KSbgr} 
T. Misumi and K. Shizuya, Phys. Rev. B {\bf 77}, 195423 (2008).


\bibitem{JHT} 
Z.~Jiang, E.~A.~Henriksen, L.~C.~Tung, Y.-J.~Wang, M.~E.~Schwartz, M.~Y. Han,
P.~Kim, and H.~L.~Stormer, Phys. Rev. Lett. {\bf 98}, 197403 (2007).


\bibitem{IWFB} A.~Iyengar, J.~Wang, H.~A.~Fertig, and L.~Brey, 
Phys. Rev. B {\bf 75}, 125430 (2007).


\bibitem{BMgr} Yu.~A.~Bychkov and G.~Martinez,
Phys. Rev. B {\bf 77}, 125417 (2008).


\bibitem{KCC} S. Viola Kusminskiy, D. K. Campbell, and A.~H.~Castro Neto, 
Europhys. Lett. {\bf 85}, 58005 (2009).


\bibitem{KScr}  K. Shizuya, Phys. Rev. B {\bf 81}, 075407 (2010);
{\bf 84}, 075409 (2011).


\bibitem{BCNM} 
Y.~Barlas, R.~C\^ot\'e, K.~Nomura, and A.~H.~ MacDonald, 
Phys. Rev. Lett. {\bf 101},
097601 (2008).


\bibitem{KSpzm}  K. Shizuya, Phys. Rev. B {\bf 79}, 165402 (2009).


\bibitem{BCLM} 
Y.~Barlas, R.~C\^ot\'e, J.~Lambert, and A.~H.~ MacDonald, 
Phys. Rev. Lett. {\bf 104}, 096802 (2010).


\bibitem{CLBM} R.~C\^ot\'e, J. Lambert,  Y. Barlas, and A. H. MacDonald,
Phys. Rev. B {\bf 82}, 035445 (2010).


\bibitem{CLPBM} R.~C\^ot\'e, W.~Luo, B.~Petrov, Y. Barlas, and A. H. MacDonald,
Phys. Rev. B {\bf 82}, 245307 (2010).


\bibitem{CFL} R.~C\^ot\'e, J.~P.~Fouquet, and W. Luo,
Phys. Rev. B {\bf 84}, 235301 (2011).


\bibitem{NL}R. Nandkishore and L. Levitov, 
Phys. Rev. B {\bf 82}, 115124 (2010);

E. V. Gorbar, V. P. Gusynin, Junji Jia, and V. A. Miransky,
Phys. Rev. B {\bf 84}, 235449 (2011).


\bibitem{Lambshift} 
W.~E.~Lamb and R.~C.~Retherford, Phys. Rev. {\bf 72}, 241 (1947).


\bibitem{KSLs}  K. Shizuya, Phys. Rev. B {\bf 86}, 045431 (2012).


\bibitem{KSLsTL}  K. Shizuya, Phys. Rev. B {\bf 87}, 085413 (2013).


\bibitem{GCP}  F. Guinea, A. H.  Castro Neto, and N. M. R. Peres, 
Phys. Rev. B {\bf 73}, 245426 (2006).


\bibitem{KA} 
M. Koshino and T. Ando, Phys. Rev. B {\bf 76}, 085425 (2007).


\bibitem{NCGP} 
J. Nilsson, A. H. Castro Neto, F. Guinea, and N. M. R. Peres, 
Phys.~Rev.~B {\bf 78}, 045405 (2008).


\bibitem{KM79} 
M. Koshino and E. McCann, Phys. Rev. B {\bf 79}, 125443 (2009).


\bibitem{BZZL} 
 W. Bao, Z. Zhao, H. Zhang, G. Liu, P. Kratz, L. Jing, J. Velasco, Jr., 
D. Smirnov, and C. N. Lau,
Phys. Rev. Lett. {\bf 105}, 246601 (2010). 


\bibitem{KEPF}  A. Kumar, W. Escoffier, J. M. Poumirol, C. Faugeras, 
D. P. Arovas, M. M. Fogler, F. Guinea, S. Roche, M. Goiran, and B. Raquet, 
Phys. Rev. Lett.  {\bf 107}, 126806 (2011).


\bibitem{TWTJ} 
T. Taychatanapat, K. Watanabe, T. Taniguchi, and P. Jarilloo-Herrero,
Nat. Phys. {\bf 7}, 621 (2011).


\bibitem{LLMC} 
C. H. Lui, Z. Li, K. F. Mak, E. Cappelluti, and T. F. Heinz, 
Nat. Phys. {\bf 7}, 944 (2011).


\bibitem{BJVL} W. Bao, L. Jing, J. Velasco Jr, Y. Lee, G. Liu, D. Tran, 
B. Standley, M. Aykol, S. B. Cronin, D. Smirnov, M. Koshino, 
E. McCann, M. Bockrath, and C. N. Lau,
Nat. Phys. {\bf 7}, 948 (2011).  


\bibitem{ZZCK} 
L. Zhang, Y. Zhang, J. Camacho, M. Khodas, and I. Zaliznyak,
Nat. Phys. {\bf 7}, 953 (2011).


\bibitem{JCST} 
S. H. Jhang, M. F. Craciun, S. Schmidmeier, S. Tokumitsu, S. Russo, M. Yamamoto, 
Y. Skourski, J. Wosnitza, S. Tarucha, J. Eroms, and C. Strunk, 
Phys. Rev. B  {\bf 84}, 161408(R) (2011).


\bibitem{HNE} 
 E. A. Henriksen, D. Nandi, and J. P. Eisenstein, 
Phys. Rev. X {\bf 2}, 011004 (2012). 


\bibitem{LVTZ} 
Y. Lee, J. Velasco, Jr, D. Tran, F. Zhang, W. Bao, L. Jing, K. Myhro,
D. Smirnov, and C.~N. Lau, Nano Lett. {\bf13}, 1627 (2013).


\bibitem{KMc81} 
M. Koshino and E. McCann, Phys. Rev. B {\bf 81}, 115315 (2010).


\bibitem{KMc83} 
M. Koshino and E. McCann, Phys. Rev. B {\bf 83}, 165443 (2011).


\bibitem{ZSMM} 
F. Zhang, B. Sahu, H. Min, and A. H. MacDonald, 
Phys. Rev. B {\bf 82}, 035409 (2010).


\bibitem{YRK} 
S. Yuan, R. Roldan and M. I. Katsnelson, Phys. Rev. B {\bf 84},
125455 (2011).


\bibitem{ZTM} 
F. Zhang, D. Tilahun, and A. H. MacDonald, Phys. Rev. B {\bf 85}, 165139 (2012).


\bibitem{fnone}  
Actually, owing to the symmetry of $ABA$ trilayers
(under the interchange of layer 1 $\leftrightarrow$ layer 3),
the full spectrum of ${\cal H}_{n}$
is a function of $|u|$, 
i.e.,  $\epsilon_{n} =\epsilon_{n}|_{\pm u}$.


\bibitem{fntwo}  
The energy difference $\Delta$ between the dimer and nondimer sites 
was left out earlier, but one can show that 
$\Delta$ induces $eh$ asymmetry but no valley asymmetry to $O(\Delta)$.

\bibitem{Toke}
Recently, the possibility of mixing of the PZM levels with levels in the velence band 
has been discussed for chiral multilayers; see C. T\"oke, arXiv:1309.5747v3. 

\end{thebibliography}
\end{document}